\documentclass{mem}
\usepackage{natbib}\usepackage{txfonts}\usepackage{balance}
\usepackage{graphicx}
\usepackage[a4paper]{hyperref}
\idline{0}{1}
\begin{document}

\title{
Non thermal emission in clusters of galaxies}

   \subtitle{}

\author{
M. \,Arnaud
          }

  \offprints{M. Arnaud}

\institute{Laboratoire AIM, DAPNIA/Service d'Astrophysique - CEA/DSM - CNRS
 - Universit\'{e} Paris Diderot, B\^{a}t. 709, CEA-Saclay, F-91191
 Gif-sur- Yvette Cedex, France
\email{monique.arnaud@cea.fr}
}

\authorrunning{M. Arnaud }

\titlerunning{Non thermal emission in Clusters of galaxies}

\abstract{I briefly review our current knowledge of the non thermal emission from galaxy clusters and  discuss future prospect with Simbol-X.  Simbol-X will map the hard X--ray  emission in clusters, determine its origin and disentangle the thermal and non-thermal components. Correlated with radio observations, the observation of the non-thermal X-ray emission, when confirmed, will allow to map both the magnetic field and the relativistic electron properties, key information to understand the origin and acceleration of relativistic particles in clusters and its impact on cluster evolution.
}
\maketitle{}

\def\etal{et al.}
\def \xmm {\hbox{\it XMM-Newton}}
\def \chandra {\hbox{\it Chandra}}
\def \sax {\hbox{\it Beppo--SAX}}
\def \xte {\hbox{\it RXTE}}
\def \lofar {\hbox{\it LOFAR}}
\def\sx  {\hbox{\it Simbol--X}}

\def\kT {{\rm k}T}
\def\keV {\rm keV}

\section{Introduction}
\begin{figure}[t]
\centering
\includegraphics[clip=true,width= 0.95\hsize]{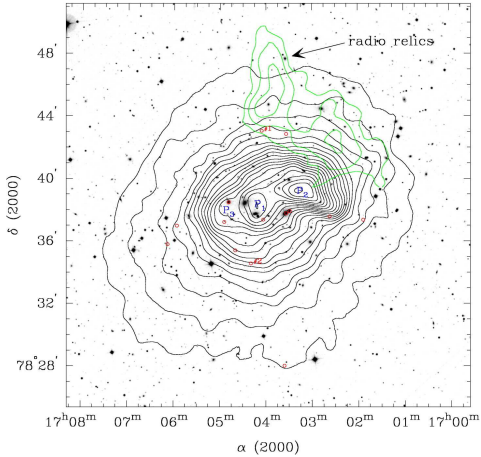}
\caption{\footnotesize Contours of the diffuse X-ray emission of A2256 as seen by \chandra\ \citep{sun02} }
\label{a2256X}
\end{figure}

\begin{figure}[ht]
\centering
\includegraphics[clip=true,width= 0.95\hsize]{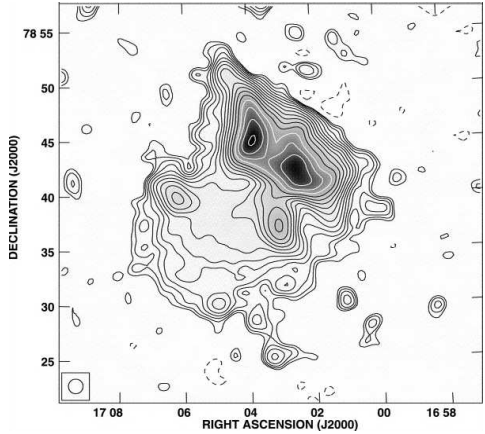}
\caption{\footnotesize VLA 1369 Mhz  emission contours A2256 \citep{cla06}.  Note the central radio halo and the  peripheral radio relic (in the North). }
\label{a2256radio}
\end{figure}

The main baryonic cluster component is a hot intergalactic gas, emitting in X-ray predominantly through thermal Bremsstrahlung (Fig.\,\ref{a2256X}). This gas is heated at typical temperatures between 2 and 10 keV during the hierarchical cluster formation: continuous accretion of surrounding matters and merger between clusters generate shocks that heat the gas to the virial temperature. In addition, large-scale magnetic field ($B$$\sim$$0.1$--$1\mu\,{\rm G}$) and relativistic electrons ($E$$\sim$$1$--$\,{\rm  few\, Gev}$) are present in the intergalactic medium, as revealed by diffuse synchrotron emission (radio halos and radio relics) observed in a fraction of galaxy clusters (Fig.~\ref{a2256radio} and Feretti, these proceedings).  These relativistic electrons can interact with the CMB photons to give inverse Compton non-thermal X-ray radiation. 

About 50 clusters with diffuse radio emission are presently known. All present evidence of recent merger events, suggesting that these events provide the energy source of the relativistic electrons. However, the origin of the relativistic particles and the acceleration processes are still poorly understood \citep[][for a review]{bru03}. A major difficulty comes from the large size of the radio halos ($\sim$Mpc). The lifetime of the electrons responsible for the emission is very short ($\sim 10^8$ years), much shorter than the diffusion time. The electrons must thus have been (re-) accelerated, or generated recently, by a mechanism acting at cluster scale. Several processes have been proposed and it is possible that all of them are important. These include injection of relativistic electrons from AGN or galactic winds, acceleration out of the thermal pool and/or re-acceleration of low energy non-thermal electrons by shocks and/or turbulence induced during merger events, continuous injection of secondary electrons by hadronic collisions of relativistic protons with the thermal gas. Similarly, the evolution of the intra-cluster magnetic field, from its primordial value, is poorly understood, although some recent numerical simulations now include it \cite[e.g.][]{dol02}.

\section{Radio and X-ray thermal emission}
The density and energy distributions of the relativistic electrons, compared to that of the thermal electrons  can help to distinguish between various models. For instance, a recent comparison of VLA radio maps and Chandra temperature maps of merging clusters suggest that radio halo electrons are mostly accelerated by turbulence rather than directly by shocks \citep{gov04}. More stringent constraints are given by radio spectral index maps. Compared with X-ray data, they provided the first confirmation that cluster mergers do supply energy to the radio halo \citep{fer04}. However, such data are presently available for only seven clusters \citep{gio93,fer04,gia05,cla06,orr07}

 Dramatic progresses in this field are expected with \lofar, as it will allow for radio multi-frequency imaging with a typical two order of magnitude gain in sensitivity, at a resolution  similar to those of X--ray satellites. \lofar\ should also detect  $\sim$1000 clusters \citep{cas06}, of which 25\% are expected to be at $z$$>$$0.3$, a unique sample for statistical studies (frequency of radio halos, correlations between X-ray and radio properties...). 
 
\section{Hard X-ray emission}
Radio observations provide important constrains on models, but not a full picture, 
the intensity of the synchrotron emission depending in a degenerate way on the strength of the magnetic field and on the relativistic electron density. The magnetic field can be estimated independently through Faraday rotation measure effects on the position angle of polarized emission from radio source viewed through the ICM  \citep[e.g.][]{gov06}.  However, the measure is limited, by nature, to a few lines of sight, i.e. this technique sparsely probes the cluster magnetic field, and parametric models of its spatial distribution have to be assumed. 

On the other hand, the Inverse Compton (IC) X-ray emission only depends on the properties of the relativistic plasma. By combining its intensity to that of the radio emission, one can break the degeneracy on the magnetic field strength and estimate the density of the relativistic electrons.  
\begin{figure}[tp]
\centering
\includegraphics[clip=true,height=0.95\hsize,angle=-90]{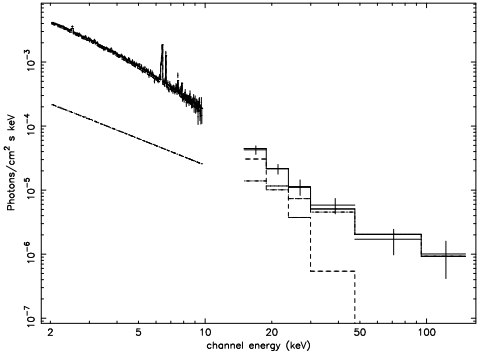}
\caption{\footnotesize Unfolded \sax\ spectrum of A2256  \citep{fus00}. The dashed line is the thermal component  ($\kT$$\sim$$7\,\keV$)  and the  dot-dashed line the  non-thermal component ($\alpha=1.4$). }
\label{a2256hxr}
\vspace{-0.5cm}
\end{figure}
For a classical power-law electron spectrum, the spectrum of the IC emission is a power law. It is expected to dominate the thermal emission only at very low energy or beyond its exponential cut-off at high energy, i.e. typically above $10\,\keV$ (Fig.~\ref{a2256hxr}).
 A significant excess of emission at high energy, with respect to an isothermal emission model, has now been detected in 13 clusters with \sax\ \citep{kaa99,fus03,fus05,fus07,nev04} or \xte\  \citep{gru02,rep02,rep03,pet06,rep06}.

\begin{figure}[tp]
\centering
\includegraphics[width=0.78\hsize]{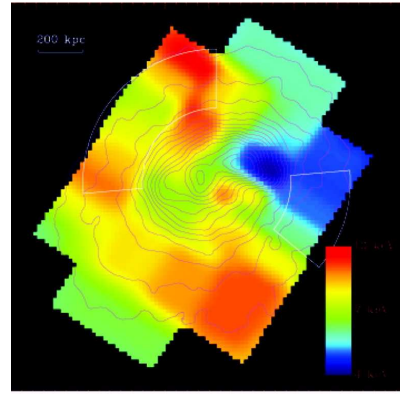}
\caption{\footnotesize Temperature map of A2256 measured with \chandra\ \citep{sun02}. X--ray emission iso-contours are overlaid. Temperature ranges from $4\,\keV$ (blue) to $10\,\keV$ (red). }
\label{a2256ktmap}
\end{figure}

 However, the significance of the excess is  low ($<$$5\,\sigma$) and  still very sensitive to possible systematic errors on the background level  \citep[][]{ros04,fus07}.  Furthermore, due to the lack of spatial information, the interpretation of the measured  {\it global} spectrum is ambiguous. It is uncertain whether the excess
   is due to IC emission or an artifact of the complex multi-temperature plasma (Fig.~\ref{a2256ktmap}) observed in non-relaxed clusters (e.g. Rephaeli \& Gruber 2003, see also Renaud \etal\ 2006).  Possible contamination by AGNs in the field of view is also an issue.

\begin{figure}[bp]
\centering
\includegraphics[clip=true,width=0.85\hsize]{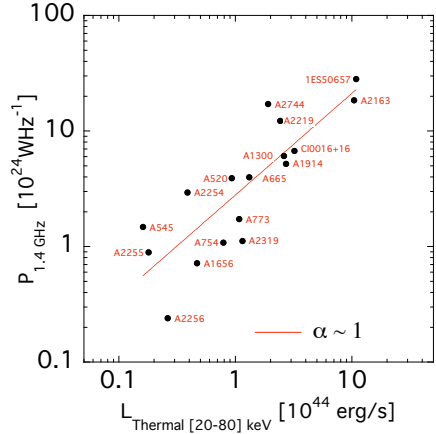}
\caption{\footnotesize Correlation between the radio power at $1.4\,{\rm Ghz}$ and the thermal hard X--ray luminosity ($[20$--$80]\,\keV$),  for known clusters with giant radio halos. Data were derived from the compilation of \citet[][]{cas06}. The dependence, $P_{\rm radio}$$\propto$$L_{\rm THXR}$, is found to be  less steep than with the soft X-ray luminosity, $P_{\rm radio}$$ \propto$$L_{\rm TSXR}^{2}$ \citep{cas06}, because the temperature increases with luminosity and thus  the $L_{\rm THXR}/L_{\rm TSXR}$ ratio.}
\label{radiohxr}
\end{figure}

 \begin{figure*}[t]
\begin{center}
\begin{minipage}[t]{0.4\hsize}
\centering
\vspace{0pt}
\includegraphics[clip=true,width=\hsize]{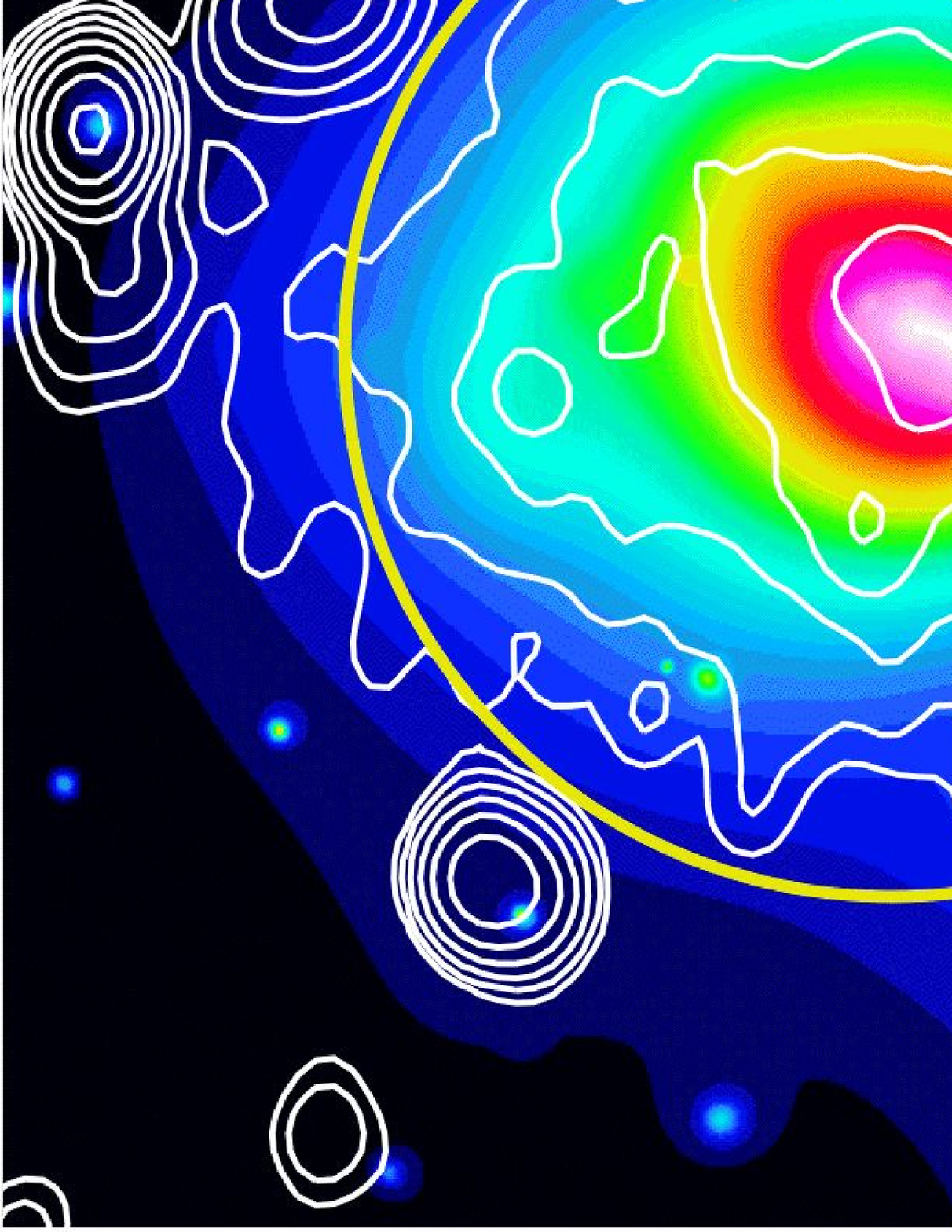}
\end{minipage}
\hspace{0.04\hsize}
\begin{minipage}[t]{0.5\hsize}
\centering
\vspace{0pt}
\includegraphics[clip=true,height=0.94\hsize, angle=-90]{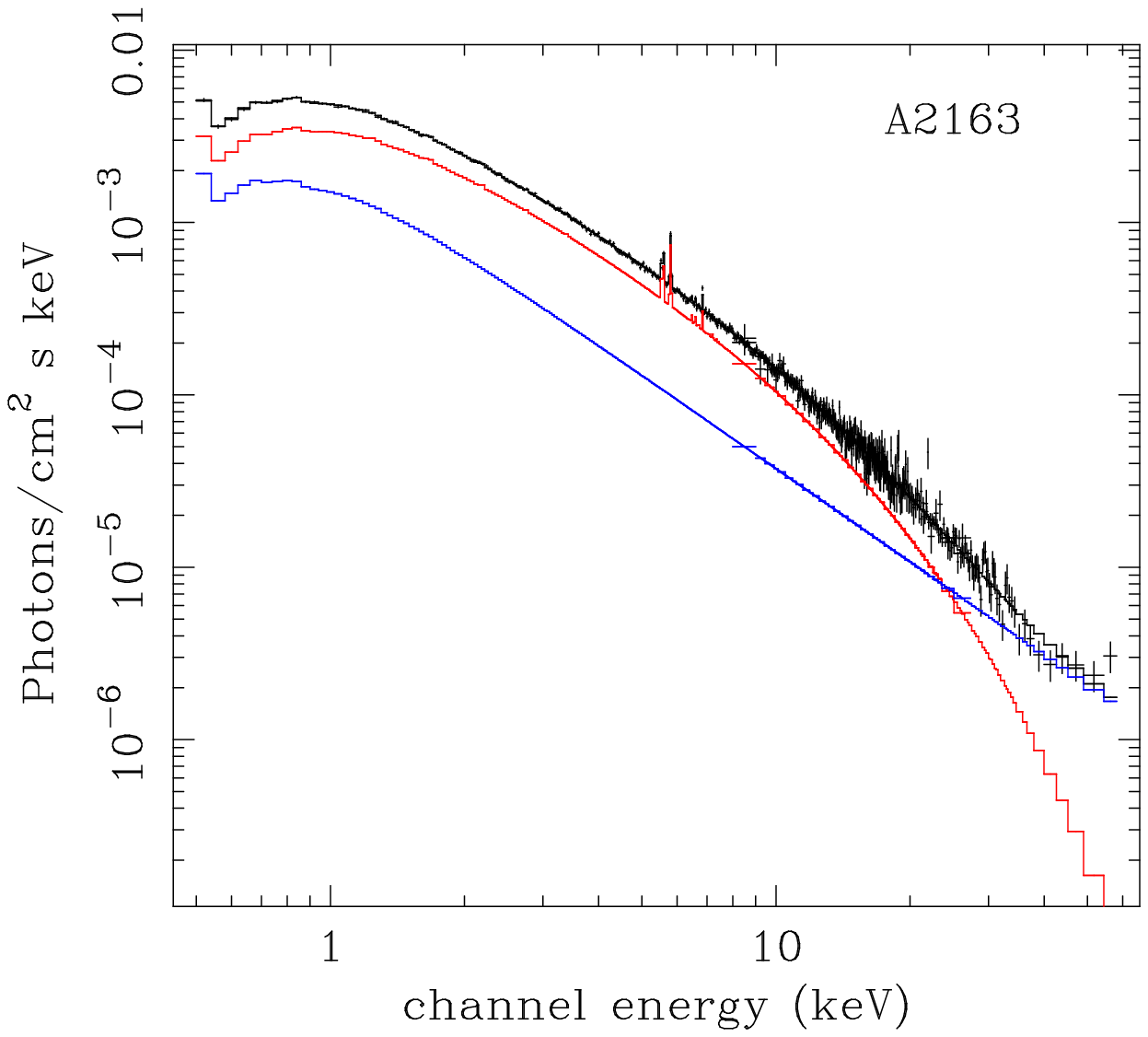}
\end{minipage}
\begin{minipage}[c]{0.4\hsize}
\includegraphics[clip=true,height=\hsize, angle=-90]{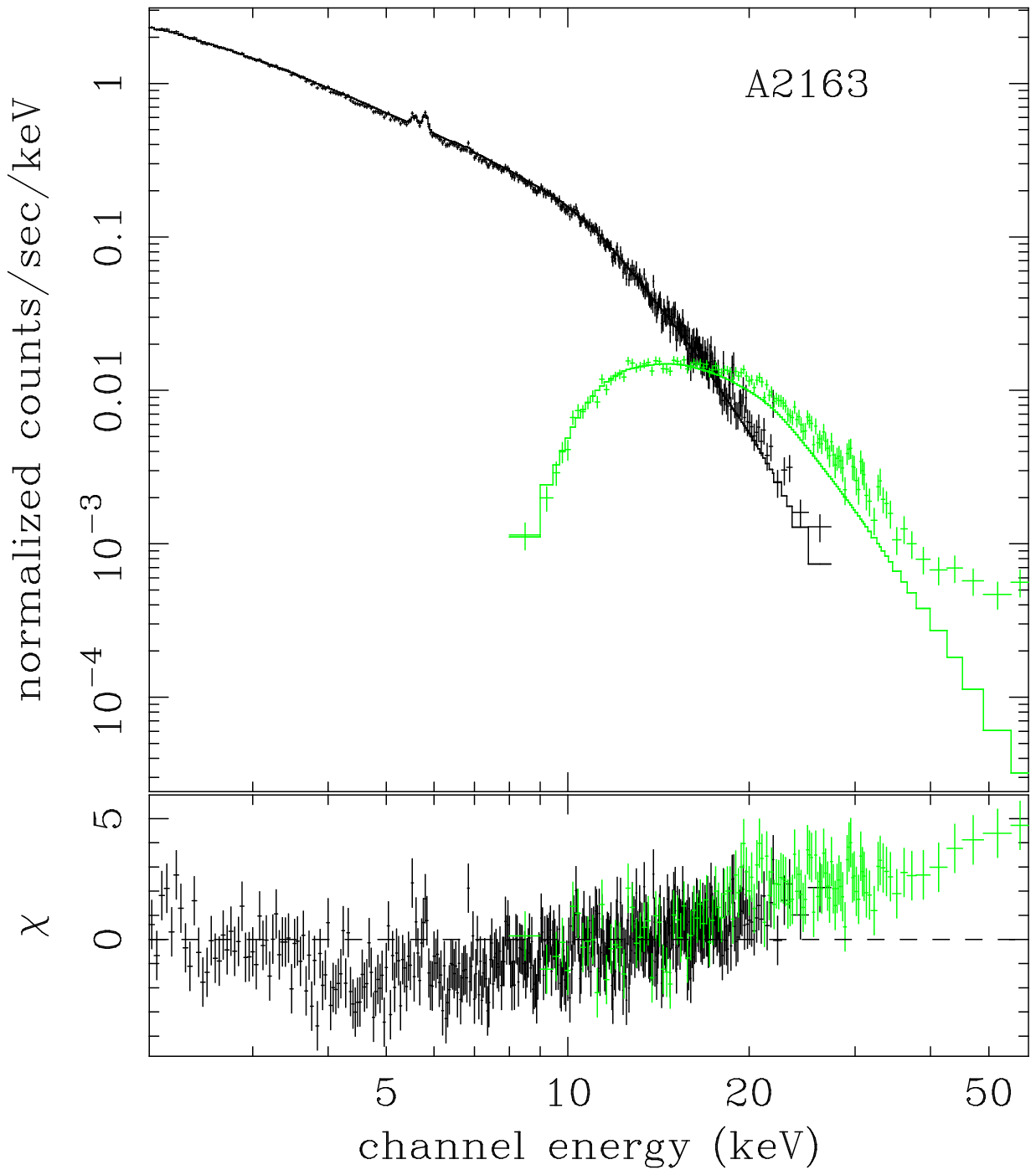}
\end{minipage}
\hspace{0.05\hsize}
\begin{minipage}[c]{0.5\hsize}
\caption{\footnotesize Simulated \sx\ spectrum of A2163 ($z$$\,=\,$$0.2$) for a 150 ksec exposure time. {\it Top-left panel}: \xmm\ image with radio contours \citep{fer01} overlaid. The circle corresponds to the  \sx\ FOV  ($11\arcmin$ diameter). {\it Top-right panel:}  Unfolded spectrum  integrated over the whole FOV. Red line: thermal component  estimated from  \xmm\ data  (mean $\kT$$\,=\,$$12\,\keV$). Blue line:  IC component, as estimated by \citet{rep06} with \xte\  ($S_{\rm IC,[20-80]\,\keV}$$\,=\,$$1.1\,10^{-11}$ergs/s/cm$^2$,  power law index of $\alpha$$\,=\,$$1.8$).  The slope and normalisation are measured with an accuracy of $\pm2\%$ and $\pm8\%$, respectively.   {\it Bottom-left panel:}  Observed spectrum with the MPD (black) and with the CZT (green) detectors. The line is the best fit isothermal model. The excess at high energy due to the IC emission is clearly detected at the $21\sigma$ confidence level, allowing its mapping (e.g $\sim 16$ sub-regions at $5\sigma$ detection level).}
\label{a2163sx}
\end{minipage}
\end{center}
\end{figure*}

\begin{figure*}[t]
\begin{center}
\begin{minipage}[t]{0.4\hsize}
\vspace{0pt}
\includegraphics[width=\hsize]{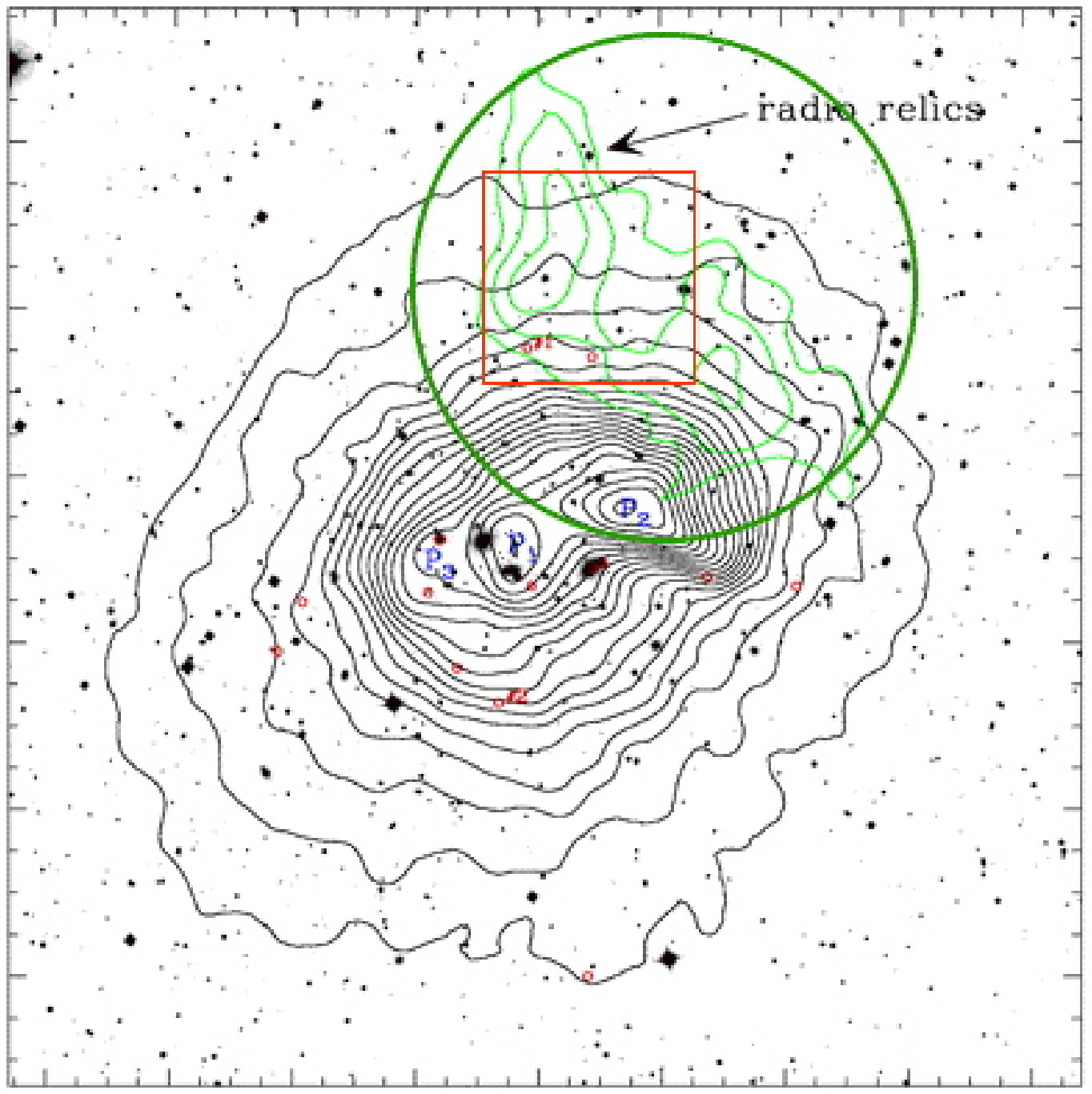}
\end{minipage}
\hspace{0.05\hsize}
\begin{minipage}[b]{0.5\hsize}
\centering
\vspace{0pt}
\includegraphics[clip=true,height=0.94\hsize, angle=-90]{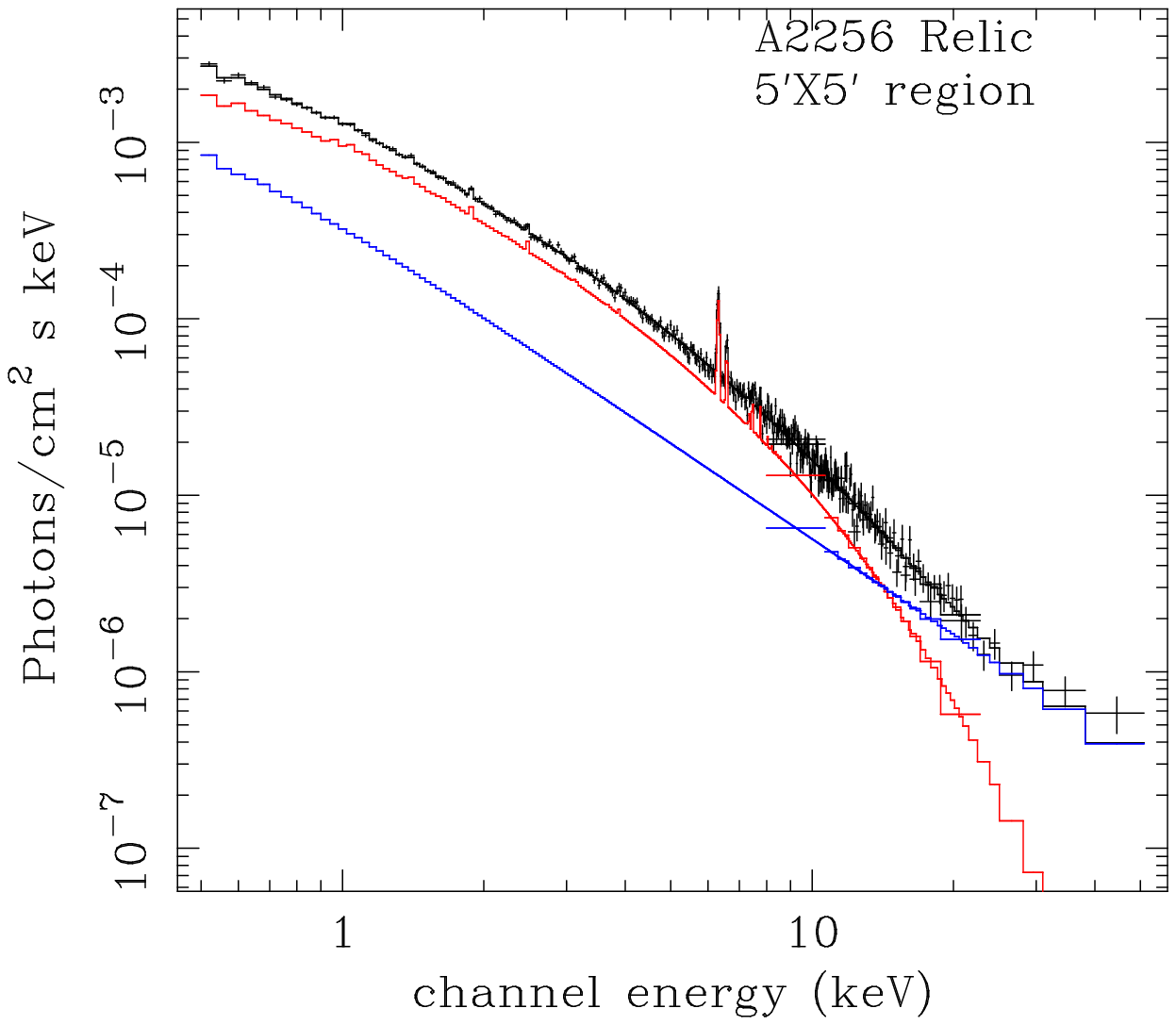}
\end{minipage}
\begin{minipage}[c]{0.4\hsize}
\includegraphics[clip=true,height=\hsize, angle=-90]{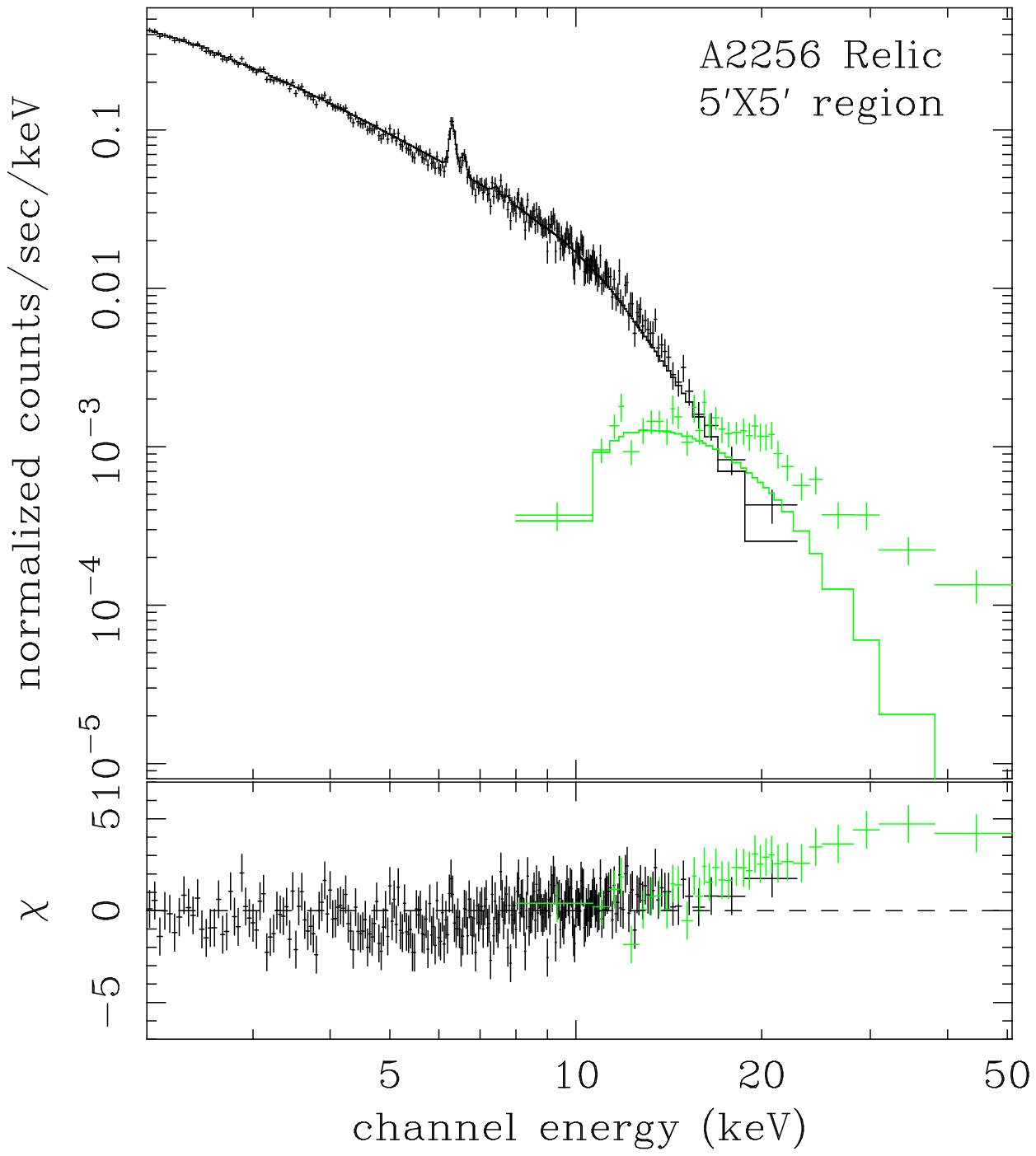}
\end{minipage}
\hspace{0.05\hsize}
\begin{minipage}[c]{0.5\hsize}
\caption{\footnotesize Same as Fig.~\ref{a2163sx} for  \sx\ observation of a $5\arcmin\times5\arcmin$ region in the A2156 radio relic (red box in top-left panel, the green circle is the  \sx\ FOV).  The thermal emission flux was estimated from {\it ROSAT} data \citep{moh99} and the temperature  $\kT$$\,=\,$$6\,\keV$ from the \chandra\ map \citep[][and Fig.~\ref{a2256ktmap}]{sun02}. Fir the IC emission, we assumed $\alpha$$\,=\,$$1.8$  and a total  flux of  $S_{\rm IC,[20-80]\,\keV}$$\,=\,$$5\,10^{-12}$ergs/s/cm$^2$,  intermediate between  \sax\ \citep{fus00} and  \xte\ value \citep{rep03} and that  it mostly originates in the relic, with $\sim\,$$1/3$ of the total flux in the  integration region. The IC emission will  be unambiguously detected, at the $12\sigma$ level (bottom-left panel) and  its slope and normalisation measured with an accuracy of $\pm6\%$ and $\pm25\%$, respectively}
\label{a2256sx}
\end{minipage}
\end{center}
\end{figure*}

\section{Measuring the cluster IC emission with \sx}

Thanks to its unique spectro-imaging capabilities up to high energies, \sx\  will allow unambiguous study of the hard X--ray (HXR)  emission from clusters. The AGNs contributions will be resolved and could be excised from the analysis. Spectra could be extracted in reasonably isothermal regions (or in regions of known thermal structure),  allowing us to disentangle the thermal and non-thermal components. 

The IC flux scales with the radio flux, redshift and magnetic field $B$, as :
\begin{equation}
S_{\rm IC} \propto S_{\rm radio} (z+1)^{2+\alpha}B^{-\alpha}~~~~ {\rm with~\alpha\sim2}. 
\end{equation}
Obviously the IC emission will be easier to detect in clusters with bright radio halos.  The radio luminosity  steeply increases with the thermal {\it soft} X-ray  luminosity, $P_{\rm radio}$$ \propto$$L_{\rm TSXR}^{2}$ \citep{cas06}, but only as $P_{\rm radio}$$\propto$$L_{\rm THXR}$ with the thermal {\it hard}  X--ray luminosity (see Fig.~\ref{radiohxr}).  In practice,  for a given $B$ value, the thermal HXR flux will be proportional  to the IC flux, and the detection of the IC emission over the thermal HXR emission will never be easy.  One can however take advantage of the different redshift dependence of the IC and radio/thermal X--ray emission, by choosing the highest z cluster for a given radio flux (Eq. 1). 

Promising \sx\ targets are thus clusters with bright radio halos at intermediate redshift ($z$$\sim$$0.2$).  Furthermore, at these redshifts, the cluster size well matches the FOV of \sx\ (Fig.~\ref{a2163sx} top panel).   An example is A2163 (Fig.~\ref{a2163sx}),  a massive cluster undergoing major merger event, which host the brightest radio halo known so far, and for which HRX excess have indeed been detected.  Note that the \lofar\ survey will make a complete sensus of clusters with bright radio halos and is expected to provide $\sim$100 suitable  targets  for \sx\ (Brunetti, these proceedings).  
\sx\ will also be particularly powerful for the study of the IC emission in radio relics, located at the cluster periphery, the thermal emission decreasing with radius (Fig.~\ref{a2256sx}).

{\it Simbol--X} will easily detect IC emission  at the level  reported by \sax\ or \xte, $S_{\rm X}$$ \sim$$10^{-11}$cgs in the $[20$--$80]\,\keV$ energy band,  with a much improved  signal to noise ratio (Fig.\,\ref{a2163sx} and Fig\,\ref{a2256sx}). However, the IC flux expected for a given radio flux is very uncertain, due to uncertainties in the magnetic field intensity (Eq. 1).  It must be noted that  estimates of  the magnetic field value  from the \sax\ or \xte\  observations of the HXR excess, assumed to be IC emission, are up to on order of magnitude lower than that estimated through Faraday depolarisation and  the actual IC flux could be significantly smaller than the values quoted above. Recent theoretical attemps significantly reduced the discrepancy, though,  \citep{bru04,col05}, assuming that the magnetic field decreases towards the cluster outskirts.  In any case,  \sx\ measurements,  combined with radio synchrotron observations,  will provide tight independent  lower limits on the $B$ value, if the IC emission is not detected, and a firm characterisation,  if it is detected at the level of previous observations.
 
 The intensity of the IC emission, when confirmed, will constrain the density of the relativistic electrons. Its spectral index, measured with high precision (Fig.~\ref{a2163sx} and Fig.~\ref{a2256sx}) will constrain the electron spectrum, complementary to the constraints provided by the radio spectrum. These measurements, giving clean access to the energetics of the relativistic particles, will be invaluable to distinguish between models. 
 Comparison between the density and energy spatial distribution of the relativistic electrons and the ICM temperature structure will provide further information on the acceleration process (e.g. via shocks or turbulence).

The study on non thermal emission with \sx\ has further cosmological implications. A better understanding of relativistic particles in clusters will provide new insights on the cluster formation process, e.g. how the energy released during merger events is redistributed between the thermal and non-thermal components. The non thermal ICM pressure will be estimated, allowing us to assess potential errors currently made when estimating total cluster mass (a key quantity when using clusters to constrain cosmological parameters) from X-ray observations and the hydrostatic equilibrium equation.

 \section{Conclusions}
 Major progresses on the IC emission in cluster are expected with \sx.  Combined with new generation radio observations, e.g.  with \lofar,  \sx\   will provide key diagnostics on the origin and acceleration of relativistic particles in clusters and its impact on cluster evolution. The science that will be done by \sx\  should also benefit  from {\it GLAST}, which will provide additional information on the protons population, and on high energy electrons accelerated at merger and accretion shocks. 

\bibliographystyle{aa}

\end{document}